\documentclass[aps,prc,preprint,groupedaddress,floatfix]{revtex4}

\usepackage{graphicx}
\usepackage{dcolumn}
\usepackage{bm}

\begin{document}

\preprint{ANU/P 1678}


\title{Some notes on the program GKINT:
Transient-field $g$-factor kinematics at intermediate energies}



\author{Andrew E.~Stuchbery}

\affiliation{Department of Nuclear Physics, Australian National
University, Canberra, ACT 0200, Australia} \email[Tel. + 61 2 6125
2097, fax + 61 2 6125 0748, email:]{andrew.stuchbery@anu.edu.au}





\date{\today}

\begin{abstract}
This report describes the computer program GKINT, which was
developed to plan, analyze and interpret the first High Velocity
Transient Field (HVTF) $g$-factor measurements on radioactive beams
produced as fast fragments. The computer program and these notes
were written in September 2004. Minor corrections and updates have
been added to these notes since then. The experiment, NSCL
experiment number 02020, {\em Excited-state configurations in
$^{38}$S and $^{40}$S through transient-field $g$-factor
measurements on fast fragments}, was performed in October 2004;
results have been published in Physical Review Letters 96, 112503
(2006).
\end{abstract}



\maketitle 
\section{Introduction}
\begin{figure}
    \resizebox{0.8\textwidth}{!}{
  \includegraphics[height=.8\textheight]{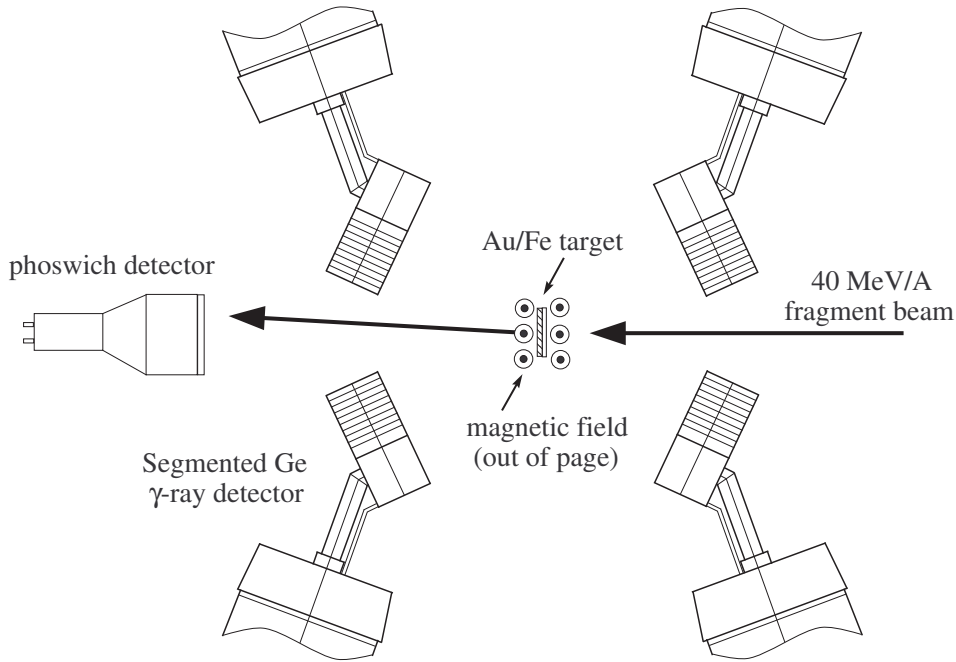}}
\caption{Schematic view of the experimental arrangement viewed from
above. Only the four SeGA detectors perpendicular to the magnetic
field axis are shown.  There are ten other SeGA detectors which are
not shown for clarity.} \label{fig:segamag}
\end{figure}

\protect \label{sect:intro}
This report contains notes on the code GKINT. Much of the code was
based on previous programs for evaluating angular correlations and
kinematics in transient-field $g$~factor measurements. These notes
therefore tend to put an emphasis on the new features that have been
introduced into GKINT. Figure~\ref{fig:segamag} shows the
experimental arrangement that the code is written to model.

\section{Outline of the program GKINT}

The code GKINT (G factor Kinematics at INTermediate energies) is
based on previous codes written at ANU to evaluate the quantities of
interest in transient-field $g$~factor measurements. The new
features include (i) the use of relativistic kinematics (ii) Coulomb
excitation calculations of cross sections and angular correlation
parameters appropriate for intermediate energies.

The target can have up to 10 layers of compound composition. One
layer has to be specified as the ferromagnetic host. In the present
version of the code, the ferromagnetic layer is assumed to be Fe.
Apart from the excitation of the 2$^+_1$ state of Fe, which is
hard-wired into the code, calculations can also be performed for any
other ferromagnetic host. (If the ferromagnet does not have $Z=26$
this part of the calculation is skipped. At present the intermediate
energy code used is not a coupled-channels code. It cannot evaluate
multiple excitation, which might be needed for a valid calculation
of the excitation of the rotational Gd isotopes, even in glancing
collisions.)

The main tasks performed by the code, in order, are:
\begin{enumerate}
\item{Read in all input data.}
\item{Trace energy loss of beam through all target layers.}
\item{Calculate projectile Coulex and angular correlation parameters as a
function of energy-loss through the target at the average scattering
angle.}
\item{Do detailed calculations of projectile Coulex cross sections,
count rates, Doppler-shift parameters, angular correlation
parameters, transit times, ion energies/velocities through each
layer, transient-field precessions, etc., taking into account both
the finite solid angle of the particle counter and the energy loss
of the beam in the target.}
\item{Evaluate angular correlations and perturbed angular correlations,
including RIV effects, at the angles of the Segmented germanium
array (SeGA).}
\item{Evaluate cross sections and angular correlations for
excitation of the Fe target layer.}
\end{enumerate}

\section{Energy loss}

\subsection{Energy loss algorithm}

Energy loss of the beam in the target layers is evaluated using the
stopping power formulation of Ziegler et al. (1985). These stopping
powers appear to be quite accurate for S beams in Au and Fe targets.
A provision is made to scale to stopping powers for each layer by an
input parameter.

Because the stopping powers are a function of energy, the energy
loss calculations have to be done by dividing the target layer into
narrow strips of width $\Delta x$. To work efficiently with the
thick foils that are used here, the energy loss algorithms were
improved by using the Runge-Kutta method.

To avoid confusing notation, we put $dE/dx = S(E)$ to denote the
energy dependence of the stopping power. If the energy of the ion as
it enters the strip $n$ is $E_n$, then the energy at the next strip
$E_{n+1}$ is
\begin{equation}
E_{n+1} = E_n - \Delta E.
\end{equation}
We have to evaluate $\Delta E$. The Runge-Kutta equations give
\begin{equation}
\Delta E = (\Delta E_1 + 2 \Delta E_2 + 2 \Delta E_3 + \Delta E_4)/6
\end{equation}
where
\begin{equation}
\Delta E_1 = S(E_n) \Delta x
\end{equation}
\begin{equation}
\Delta E_2 = S(E_n-\Delta E_1/2) \Delta x
\end{equation}
\begin{equation}
\Delta E_3 = S(E_n-\Delta E_2/2) \Delta x
\end{equation}
\begin{equation}
\Delta E_4 = S(E_n- \Delta E_3) \Delta x
\end{equation}

The Runge-Kutta method gives an exact solution (but for rounding
errors) when $S(E)$ can be represented by a polynomial of degree not
greater than four. Convergence is tested by repeating the
calculation with double the number of steps. The energy loss
converges within a few keV for targets several hundred mg/cm$^2$
thick when a few hundred steps are used. There can be a reduction in
the accuracy if the ion almost stops, but such cases must be avoided
in an experiment anyway.

\subsection{Energy loss calculations in GKINT}

The first part of the calculation is to evaluate the energy loss of
the beam through the target layers. The beam energy at the front of
each target layer is determined and the transit time through the
layer is evaluated. The energy with which the ion leaves the target
is also printed.

For the subsequent integrations, each target layer is then
subdivided into a number of strips of equal depth, and the energy of
the beam at each strip is evaluated and stored. A survey of the
Coulomb excitation cross sections and angular correlation parameters
($a_2$ and $a_4$) as a function of depth is made, assuming that the
beam scatters at the average detection angle, $\bar{\theta} =
(\theta_{\rm max} + \theta_{\rm min})/2$.

In the case of intermediate-energy fragments scattering at small
angles, the difference between the incident and scattered beam
energies is very small. However it is evaluated precisely, and the
subsequent energy loss of the scattered beam ion along its
trajectory toward the particle counter is evaluated as a function of
depth through the target, as required.

\section{Transient-field precessions}

Transient-field precessions are evaluated for ions that traverse the
ferromagnetic layer of the target using the routine ROTN. The
primary purpose of this routine is to evaluate
\begin{equation}
\phi(\tau) = \Delta \theta/g  = -  \frac{\mu _N}{\hbar} \int ^{T _e}
_{T _i } B_{\rm tr} (t) {\rm e} ^{-t/\tau} \, {\rm d}t,
\label{eq:phi1}
\end{equation}
where $g$ and $\tau$ are the $g$~factor and lifetime of the excited
state. $B_{\rm tr}$ is the transient-field strength, which varies
with time as the ion slows within the ferromagnet.

The integral over time is rewritten as an integral over energy using
the non-relativistic relations
\begin{equation}
dt = dx/v = dx /\sqrt{2E/M} = dE \sqrt{M/2E} /(dE/dx)
\end{equation}
where $E$ is the energy and $M$ is the mass of the ion. The use of
the non relativistic formula for energy is not expected to be too
unrealistic because the TF strength is small until the ion velocity
falls to around $0.15c$.

In the routine ROTN the integral over the energy range is performed
using Simpson's rule. Along with the precession angle, ROTN
evaluates a number of other quantities that are often considered
useful in relation to the Transient Field.

The effective time for which the ion experienced the TF during its
flight through the ferromagnetic foil layer must be evaluated first.
From this, one may derive the average ion velocity and the average
TF strength.

If $t_i$ and $t_e$ are the times at which the probe ion entered into
and exited from the ferromagnetic foil, and $\tau$ is the mean life
of the probe state, the effective interaction time is
\begin{equation}
t_{\rm eff} = \tau [ \exp(-t_i / \tau) - \exp(-t_e / \tau)].
\end{equation}
When $\tau \gg t_i, t_e$,
\begin{equation}
t_{\rm eff} = t_i - t_e = t_{fe},
\end{equation}
where $t_{fe}$ is the time the probe ion takes to traverse the
ferromagnetic layer. The average velocity of the ion whilst in
transit through the ferromagnet is
\begin{equation}
\langle v/v_0 \rangle = [ \int_{t_i}^{t_e} (v/v_0) \exp(-t/\tau)
dt]/t_{\rm eff} ,
\end{equation}
where $v_0 = c/137$ is the Bohr velocity. Again, when the mean life
is much longer than the transit time,
\begin{equation}
\langle v/v_0 \rangle = L/[v_0 t_{fe}],
\end{equation}
where $L$ is the thickness of the ferromagnetic foil. If $\phi$ is
expressed in milliradians and the interaction time in ps, the
average TF strength in kTesla is
\begin{equation}
\langle B_{TF} \rangle = 0.0209 \phi / t_{\rm eff}.
\end{equation}

\section{Coulex calculations and particle kinematics}

\subsection{Modifications to COULINT}

Coulex cross sections are evaluated using routines from COULINT,
written by C.A. Bertulani, see Phys. Rev. C 68, 044609 (2003). A few
modifications were needed for the present application.

Firstly, the COULINT routines were modified to return the CM
differential cross section for a specified energy and scattering
angle. Both the Rutherford and Coulomb excitation cross sections are
returned. The code was also modified to return the statistical
tensor for the angular correlation in the projectile frame, at the
specified energy and particle scattering angle. Since there is some
difference in the terminology, and in the definitions of the
statistical tensors, the code works with $a_k$ coefficients, which
are the coefficients that occur in the familiar Legendre polynomial
expansion of the angular correlation:
\begin{equation}\protect\label{eq:wtheta}
W(\theta_\gamma) = 1 + a_2 P_2(\cos \theta_\gamma) + a_4 P_4(\cos
\theta_\gamma)
\end{equation}

The centre-of-mass to Lab transformations were evaluated with
relativistic formulae:
\begin{equation}
\tan \theta_{\rm Lab} = \frac{1}{\gamma_2} \cdot \frac{\sin
\theta_{\rm CM}}{(\cos \theta_{\rm CM} + M_p \gamma_1/ M_t \gamma_2
)}
\end{equation}
and
\begin{equation}
\cos \theta_{\rm CM} = \frac{1}{(1 + \gamma_2^2 \tan^2 \theta_{\rm
Lab})} \left [-\frac{M_p}{M_t} \gamma_1 \gamma_2 \tan^2 \theta_{\rm
Lab} \pm \left (1 - \frac{M_p^2 - M_t^2}{M_t^2} \tan^2 \theta_{\rm
Lab} \right )^{1/2} \right ]
\end{equation}
where the positive sign is the only one relevant in the present
work. The notation used here drops the prime that in many references
is attached to the factors $\gamma_i$. To define these gamma factors
it is convenient to introduce $\rho = M_p/M_t$. As usual we have
\begin{equation}
\gamma = (1 - v^2/c^2)^{-1/2} = 1 + T/m_0c^2
\end{equation}
where $T$ is the beam energy in MeV per nucleon and $m_0c^2 =
931.494$~MeV. In terms of $\rho$ and $\gamma$ we have
\begin{equation}
\gamma_1 = \frac{\gamma + \rho}{(1 + 2 \rho \gamma + \rho^2)^{1/2}}
\end{equation}
and
\begin{equation}
\gamma_2 = \frac{1 + \gamma \rho}{(1 + 2 \rho \gamma +
\rho^2)^{1/2}}
\end{equation}

The lab cross sections were evaluated as
\begin{equation}
\frac{d \sigma _{\rm Lab} }{d \Omega _{\rm Lab} } = \frac{d \sigma
_{\rm CM} }{d \Omega _{\rm CM} } \cdot \frac{d \Omega _{\rm CM} }{d
\Omega _{\rm Lab} }
\end{equation}

The solid angle ratio, evaluated non-relativistically in terms of
the particle scattering angles, is
\begin{equation}
\frac{(d \sigma / d \Omega)_{\rm CM} }{(d \sigma / d \Omega )_{\rm
Lab} } = \frac{d \Omega _{\rm Lab} }{d \Omega _{\rm CM}}  = \left (
\frac{\sin \theta_{\rm Lab}}{\sin \theta_{\rm CM}} \right ) ^2
|\cos(\theta_{\rm CM} - \theta_{\rm Lab})|
\end{equation}

The relativistic formula is
\begin{equation}
\frac{d \Omega _{\rm Lab} }{d \Omega _{\rm CM}}  = \frac{\gamma_2[1
+  (M_p \gamma_1/ M_t \gamma_2)\cos \theta_{\rm CM} ]}{[\sin^2
\theta_{\rm CM} + \gamma_2^2 (\cos \theta_{\rm CM} + M_p \gamma_1/
M_t \gamma_2)^2 ]^{3/2}}
\end{equation}

The code uses the proper relativistic evaluation, although the
relativistic formulation differs from the non-relativistic one by a
negligible factor in many cases.

\subsection{GKINT description: cross sections}

\begin{figure}
    \resizebox{0.8\textwidth}{!}{
  \includegraphics[height=.8\textheight]{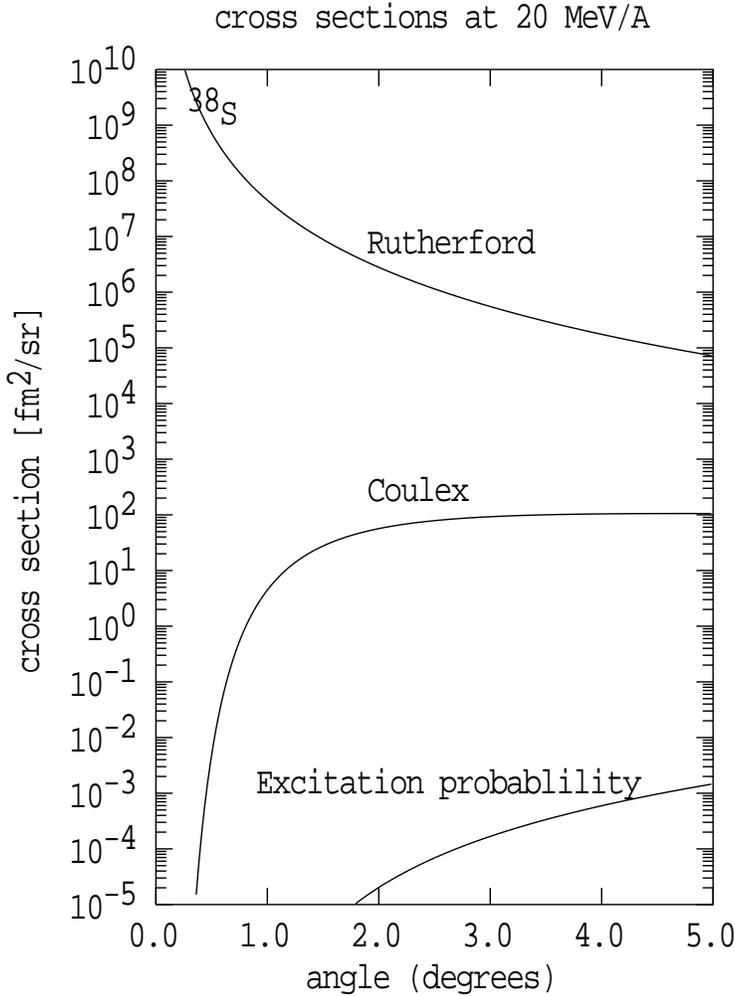}}
\caption{Cross sections as a function of scattering angle for 20
MeV$/A$ $^{38}$S on $^{197}$Au.} \label{fig:x20}
\end{figure}

\begin{figure}
    \resizebox{0.8\textwidth}{!}{
  \includegraphics[height=.8\textheight]{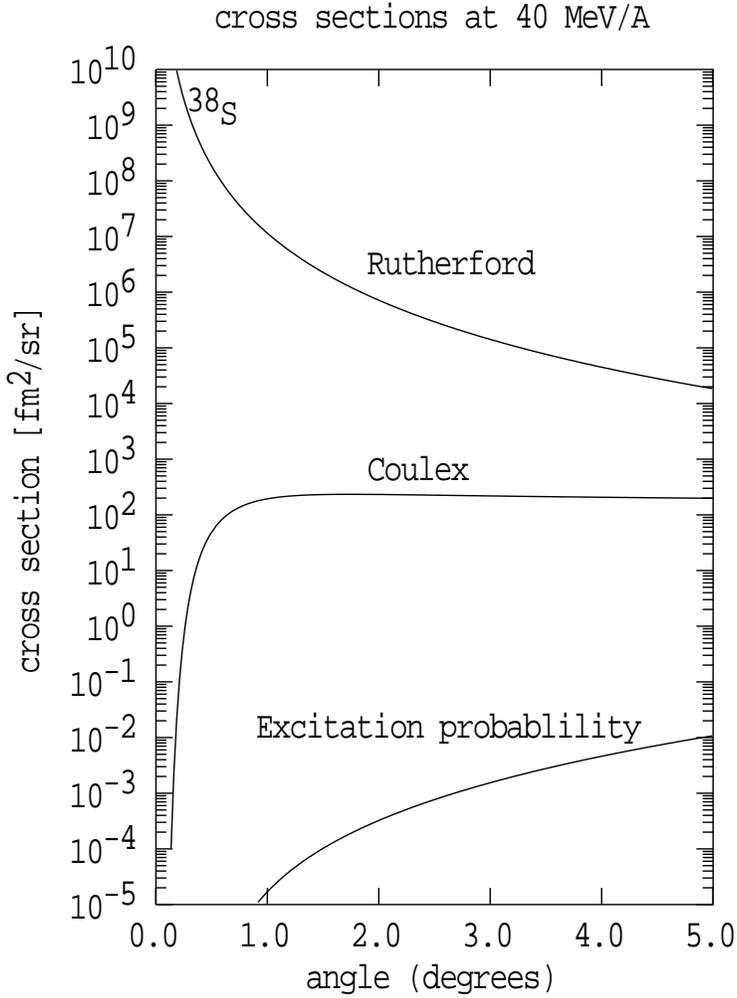}}
\caption{Cross sections as a function of scattering angle for 40
MeV$/A$ $^{38}$S on $^{197}$Au.} \label{fig:x40}
\end{figure}

Figures \ref{fig:x20} and \ref{fig:x40} show the variations in the
projectile Coulomb excitation and Rutherford cross sections as a
function of scattering angle for $^{38}$S at 20 and 40 MeV$/A$,
respectively, incident on $^{197}$Au. The figures also show the
excitation probability, $P_{i \rightarrow f}$, which is
\begin{equation}
P_{i \rightarrow f} = \frac{(d\sigma/d\Omega)_{i \rightarrow f}}
{(d\sigma/d\Omega)_{\rm Rutherford}}
\end{equation}
Note that the Rutherford cross section goes to infinity as the
scattering angle approaches zero. The Coulex cross section remains
well behaved, however, because $P_{i \rightarrow f}$  approaches
zero faster than the Rutherford cross section diverges.

GKINT performs integrals over the solid angle of the particle
detector and over the depth of the target:
\begin{equation}
\langle \sigma \cdot t_{\rm tgt} \rangle = 2 \pi \int_{0}^{t_{\rm
tgt}} \int_{\theta_{p}^{\rm min}}^{\theta_{p}^{\rm max}} \frac{d
\sigma }{d \Omega }(\theta_p , E[z])  \; \sin \theta_p d \theta_p d
z,
\end{equation}
where the scattering angles and the cross section are in the lab
frame, and $z$ denotes the depth through the target. With the cross
sections in fm$^2$ and target thickness in mg/cm$^2$, the count rate
$R$ in Hz is
\begin{eqnarray}
R &=& N_{\rm beam} \langle \sigma \cdot t_{\rm tgt} \rangle
\frac{N_A}{A}\\
 &=& (6.023 \times 10^{-6}) N_{\rm beam}
\frac{\langle \sigma \cdot t_{\rm tgt} \rangle}{A},
\end{eqnarray}
where $N_{\rm beam}$ is the number of beam ions/sec, $N_A$ is
Avagadro's number, and $A$ is the atomic mass number. The count
rates for Coulomb excitation and Rutherford scattering as registered
by the particle counter are printed for $N_{\rm beam}=1000$. If the
Rutherford cross section is in the region where it diverges and thus
implies $R >  N_{\rm beam}$, which is clearly unphysical, the code
puts $R = N_{\rm beam}$ and prints a warning.

\subsection{GKINT description: angular correlation coefficients}

\begin{figure}
    \resizebox{0.8\textwidth}{!}{
  \includegraphics[height=.8\textheight]{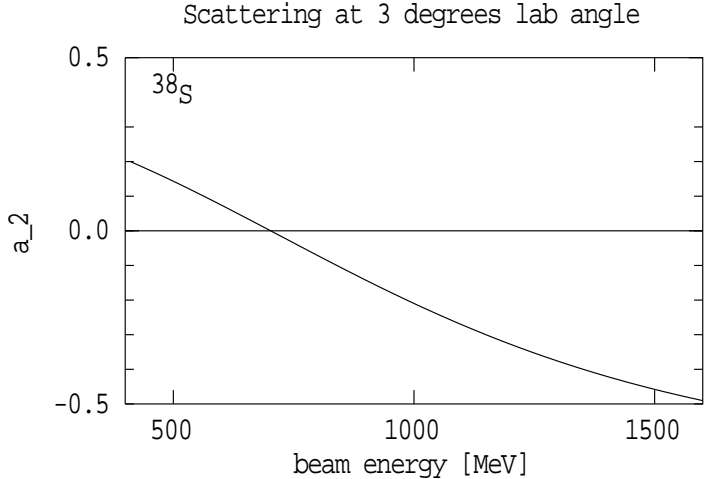}}
\caption{$a_2$ angular correlation coefficient as a function of beam
energy at a lab scattering angle of 3$^\circ$ for $^{38}$S on
$^{197}$Au.} \label{fig:a2plt}
\end{figure}

\begin{figure}
    \resizebox{0.8\textwidth}{!}{
  \includegraphics[height=.8\textheight]{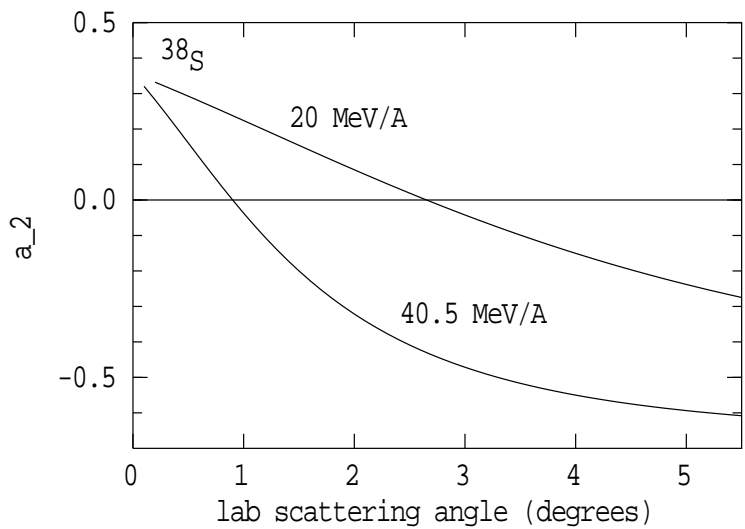}}
\caption{$a_2$ angular correlation coefficient as a function of lab
scattering angle for $^{38}$S on $^{197}$Au.} \label{fig:a2thet}
\end{figure}

The spin alignment in intermediate energy Coulomb excitation depends
on both the beam energy and the scattering angle. As can be seen in
Fig.~\ref{fig:a2plt}, the alignment decreases as the beam energy
decreases, eventually changing sign. Similar behavior is evident
when $a_2$ is plotted as a function of the scattering angle,
Fig.~\ref{fig:a2thet}. Again the alignment decreases with scattering
angle, eventually changing sign.

For each layer of the target, the average $a_k$ coefficients are
evaluated as
\begin{equation}\protect\label{eq:ak}
\langle a_k \rangle = \frac{2 \pi}{\langle \sigma \cdot t_{\rm tgt}
\rangle} \int_{0}^{t_{\rm tgt}} \int_{\theta_{p}^{\rm
min}}^{\theta_{p}^{\rm max}} a_k(\theta_p, E[z]) \frac{d \sigma }{d
\Omega }(\theta_p , E[z])  \; \sin \theta_p d \theta_p d z.
\end{equation}
The integrals are performed numerically using Simpson's rule.

Similar integrals are performed to evaluate the average energy of
excitation, the average transient-field precessions, etc. In each
case the quantity to be averaged replaces $a_k$ in an expression of
the same form as Eq.~(\ref{eq:ak}).

\section{Angular correlation formalism}

The theoretical expression for the unperturbed angular correlation
after Coulomb excitation in general can be written as
\begin{equation}
W(\theta_\gamma, \phi_\gamma) = \sum_{k,q} \sqrt{(2k+1)}
\rho_{kq}(\theta_p,\phi_p) F_k G_k Q_k
D^{k*}_{q0}(\phi_\gamma,\theta_\gamma,0) ,
\end{equation}
where $\rho_{kq}(\theta_p,\phi_p)$ is the statistical tensor which
defines the spin alignment of the initial state, and depends on the
particle scattering angles $(\theta_p,\phi_p)$, $F_k$ represents the
usual $F$-coefficients for the $\gamma$-ray transition, $G_k$ is the
vacuum deorientation attenuation factor, $Q_k$ is the attenuation
factor for the finite size of the $\gamma$-ray detector and
$D^k_{q0}(\phi_\gamma,\theta_\gamma,0)$ is the rotation matrix,
which depends on the $\gamma$-ray detection angles
$(\theta_\gamma,\phi_\gamma)$. The Coulex calculation uses a
right-handed co-ordinate frame in which the beam is along the
positive $z$-axis. In the applications of interest $k =0,2,4$.

Since we are interested in the case where the particle detector has
azimuthal symmetry, the statistical tensors $\rho_{kq}$ depend only
on $\theta_p$ and are non-zero only when $q=0$:
\begin{equation}
\rho_{kq}(\theta_p,\phi_p) = \delta_{q,0} \rho_{k0}(\theta_p)
\end{equation}

The program GKINT integrates the Coulex cross sections over the
opening angle of the particle detector and evaluates the $a_k$
coefficients defined as
\begin{equation}
a_k = \sqrt{(2k+1)} \rho_{k0} F_k
\end{equation}

In the experiments a magnetic field will be applied in the vertical
direction, along the $x$-axis or $\phi=0$ direction. We define FIELD
UP as being when the external field points UP along the positive
$x$-axis. The transient-field precession is given by
\begin{equation}
\Delta \theta = g \phi = - g \frac{\mu _N}{\hbar} \int ^{T _e} _{T
_i } B_{\rm tr} (t) {\rm e} ^{-t/\tau} \, {\rm d}t, \label{eq:phi}
\end{equation}
where $g$ is the nuclear $g$ factor, $\tau$ is the meanlife of the
level, and $T_i$ and $T_e$ are the times at which the ion enters
into and exits from the ferromagnetic foil. (This expression
corresponds to the precession information in the whole gamma-ray
lineshape, including decays which occur while the nucleus is still
in motion through the ferromagnet. The code also evaluates the
precession for those nuclei that decay in vacuum after leaving the
ferromagnetic layer of the target.) The important point to note here
is that the precession angle, $\Delta \theta$, is {\em negative} for
a positive field (i.e. field UP) and a positive $g$~factor.

Now consider that the nucleus undergoes a transient-field precession
of $\pm \Delta \theta$ as the external field is alternatively
flipped up and down. We set aside the question of whether the
positive precession corresponds to field-up or field-down, since we
don't know the sign of the $g$~factor a priori. The statistical
tensor is then modified by the precession about the field axis,
according to
\begin{eqnarray}
\rho_{kq}(\pm \Delta \theta) &=& \sum_{Q} \rho_{kQ} \delta_{Q,0}
D^k_{qQ}(\pi/2, \pm \Delta \theta, -\pi/2)\\
&=&\rho_{k0} D^k_{q0}(\pi/2, \pm \Delta \theta, 0)\\
&=&\rho_{k0} D^k_{q0}(\pm \pi/2,  |\Delta \theta|, 0)\\
&=&\rho_{k0} \sqrt{\frac{4 \pi}{(2k+1)}} Y^k_{q}(|\Delta \theta|,\pm
\pi/2)
\end{eqnarray}

In GKINT it is convenient to work with spherical harmonics rather
than the rotation matrices. To get the final form of the perturbed
angular correlation we need to use
\begin{equation}
D^k_{q0}(\alpha, \beta, \gamma) = \sqrt{\frac{4 \pi}{(2k+1)}}
Y^k_{q}(\beta,\alpha)
\end{equation}
and
\begin{equation}
Y^{k*}_{q}(\beta,\alpha) = (-1)^q Y^k_{- q}(\beta,\alpha)
\end{equation}
Note that, depending on the implementation in the computer routine,
putting a negative $\beta$ value in the $D$ matrix, and/or the
spherical harmonic, may not be properly interpreted. We will
therefore avoid doing it in the code.

We then get the final expression for the perturbed angular
correlation which GKINT uses:
\begin{equation}
W(\theta_\gamma, \phi_\gamma, \pm \Delta \theta) = 4 \pi \sum_{k,q}
\frac{a_k Q_k G_k}{(2k+1)} (-1)^q Y^k_{q}(|\Delta \theta|,\pm \pi/2)
Y^k_{- q}(\theta_\gamma,\phi_\gamma)
\end{equation}

The evaluation of the vacuum deorientation parameters, $G_k$, is
discussed below. At present the gamma detector solid angle factors
$Q_k$ are all set to unity. We will need to formulate a procedure to
evaluate these in future (perhaps using GEANT). However the
detectors are sufficiently far from the target that $Q_k = 1$ is a
reasonable first approximation.

\section{Lorentz correction for $\gamma$ rays}

Because we have azimuthal symmetry, the Lorentz correction in this
case is quite straight-forward. The basic formulae, which have been
written in many publications, are reviewed in the first subsection.
In the second subsection the method of treating the perturbed
angular correlation is discussed. I'm not aware of any place in the
literature where this situation is discussed in any detail.

\subsection{Formulae for angular correlations}

The angle between the beam axis and the direction of emission of the
photon is affected if the nucleus is in motion rather than at rest.
The relativistic velocity addition formulae  can be used to show
that the angles of emission in the laboratory, $(\theta_{\rm lab},
\phi_{\rm lab})$, are related to those in the rest frame of the
nucleus, $(\theta_{\rm nuc},\phi_{\rm nuc})$, by the expressions
\begin{equation}\label{eq:thtrans}
\cos \theta_{\rm lab} = \frac{\cos \theta_{\rm nuc} + \beta} {1+
\beta \cos \theta_{\rm nuc}} ,
\end{equation}
and
\begin{equation}
\phi_{\rm lab} = \phi_{\rm nuc} ,
\end{equation}
where $\beta = v/c$ is the velocity of the nucleus along
$\theta_{\rm lab} = \theta_{\rm nuc} =0$. It follows that an element
of solid angle in the rest frame is related to an element of solid
angle in the laboratory frame by
\begin{equation}
{\rm d}\Omega_{\rm lab} = \frac{1 - \beta^2}{(1 + \beta \cos
\theta_{\rm nuc})^2}{\rm d}\Omega_{\rm nuc}.
\end{equation}
To obtain the correlation in the laboratory frame, $W_{\rm
lab}(\theta_{\rm lab})$, from that in the rest frame of the nucleus,
$W_{\rm nuc}(\theta_{\rm nuc})$, requires both transformation of the
laboratory angle to the equivalent angle in the rest frame of the
nucleus using Eq.~(\ref{eq:thtrans}), and multiplication by the
appropriate solid-angle ratio so that the $\gamma$-ray flux emitted
into $4 \pi$ of solid angle is conserved, i.e.
\begin{equation}
\int_{4 \pi} W_{\rm lab}(\theta_{\rm lab}) {\rm d}\Omega_{\rm lab} =
\int_{4 \pi} W_{\rm nuc}(\theta_{\rm nuc}) {\rm d}\Omega_{\rm nuc}.
\end{equation}
This condition is satisfied if
\begin{equation}
W_{\rm lab}(\theta_{\rm lab}) = W_{\rm nuc}(\theta_{\rm nuc})
\frac{{\rm d}\Omega_{\rm nuc}}{{\rm d}\Omega_{\rm lab}}.
\end{equation}

\subsection{Formulae for perturbed angular correlations}

The precession of the nucleus takes place in the frame of the
nucleus (obviously!). However we measure the perturbed angular
correlation in the Lab frame.

The standard technique for measuring small transient-field
precessions in a way that minimizes possible systematic errors is to
form a double ratio of $\gamma$-ray intensities for each
complementary pair of $\gamma$-ray detectors placed at angles $\pm
\theta_{\gamma}$ with respect to the beam direction:
\begin{equation}
\rho = \sqrt{\,\frac{{\rm N}(+\theta_\gamma \!\!\uparrow)} {{\rm
N}(+\theta_\gamma \!\!\downarrow)}\: \frac{{\rm N}(-\theta_\gamma
\!\!\downarrow)} {{\rm N}(-\theta_\gamma \!\!\uparrow)}} \;\;\; ,
\end{equation}
where N$(\pm \theta_\gamma \uparrow,\downarrow$) denote the numbers
of counts in the photopeak of the transition of interest as
registered in the detectors at $\pm \theta_\gamma$ for the
alternative field directions ($\uparrow$ and $\downarrow$).

The effect of the transient field is to rotate the angular
correlation by a relatively small angle around the direction of the
applied magnetic field. Thus

\begin{equation}
W(\theta_{\gamma} \uparrow) = W(\theta_{\gamma} - \Delta \theta)
\simeq W(\theta_{\gamma} ) - \Delta \theta \left . \frac{{\rm d}
W}{{\rm d} \theta} \right | _{\theta_\gamma} .
\end{equation}

In the case where there are no (or negligible) Lorentz effects, the
precession angle $\Delta \theta$ is determined from
\begin{equation}
 \Delta\theta = \frac{\varepsilon}{ S} \;\;\; ,
\label{dtheta}
\end{equation}
where
\begin{equation}
\label{epsilon}
 \varepsilon = \frac{1-\rho}{1+\rho} \;\;\; ,
\end{equation}
and $S$, the logarithmic derivative of the angular correlation at
the angle $\theta_\gamma$, is
\begin{equation} \label{s}
S = \left . \frac{1}{W} \frac{{\rm d} W}{{\rm d} \theta} \right |
_{\theta_\gamma} \;\;\; .
\end{equation}

To see how these expressions must be modified to account for Lorentz
effects, it is convenient to note that $ \varepsilon$ is formally
equivalent to
\begin{equation}
 \varepsilon = \frac{W(+\theta_\gamma\!\! \downarrow) -
 W(+\theta_\gamma \!\! \uparrow)}
 {W(+\theta_\gamma \!\! \downarrow) + W(+\theta_\gamma \!\! \uparrow)}
 .
\end{equation}

The perturbed angular correlation in the lab frame is given by
\begin{equation}
W_{\rm lab}(\theta_{\rm lab} \uparrow) = W_{\rm nuc}(\theta_{\rm
nuc} \uparrow) \frac{{\rm d}\Omega_{\rm nuc}}{{\rm d}\Omega_{\rm
lab}} = W_{\rm nuc}(\theta_{\rm nuc} - \Delta \theta) \frac{{\rm
d}\Omega_{\rm nuc}}{{\rm d}\Omega_{\rm lab}}.
\end{equation}
Hence
\begin{equation}
 \varepsilon_{\rm lab} = \frac{W(\theta_{\rm nuc} \!\! \downarrow) -
 W(\theta_{\rm nuc} \!\! \uparrow)}
 {W(\theta_{\rm nuc} \!\! \downarrow) + W(\theta_{\rm nuc} \!\! \uparrow)}
=  \varepsilon_{\rm nuc}.
\end{equation}
In the case of the Lorentz boost, the expression for $\Delta \theta$
is
\begin{equation}
 \Delta\theta = \frac{\varepsilon_{\rm lab}}{ S_{\rm nuc}} \;\;\; .
\end{equation}
Thus the only change in the analysis procedure is that $S$ must be
evaluated at the appropriate angle in the rest frame of the nucleus,
which corresponds to the laboratory detection angle.

Finally, it is worth noting that if $W(\theta)$ has the form of
Eq.~(\ref{eq:wtheta}), then
\begin{equation}
S(\theta) = -(a_2 P_{21}(\cos \theta) + a_4 P_{41}(\cos
\theta))/W(\theta)
\end{equation}
where $P_{21}$ and $P_{41}$ are associated Legendre polynomials.

\section{Recoil in Vacuum Attenuation Factors}

The $G_k$ coefficients are evaluated as follows, based on
experimental data and experience at ANU.

For the experiments of interest, with ions emerging into vacuum with
velocities near the $K$-shell electron velocity, the vacuum
deorientation is dominated by the fields at the nucleus from the
ground-state of the H-like electron configuration, i.e. $1s$ fields.
At the lower velocities of interest $2s$ fields also contribute. In
this description, we will begin by assuming that the $1s$ fields are
dominant and then introduce the $2s$ contribution at the end. The
code uses both.

Neglecting the relativistic correction (which is small for $Z=16$),
the field at the nucleus from the $1s$ electron of an H-like ion is
\begin{equation}
B_{1s} = 16.7 Z^3 \;\;\; {\rm Tesla.}
\end{equation}
This free-ion field causes the nuclear spin $I$ to precess with a
Larmor frequency of
\begin{equation}
\omega = g \frac{\mu _N}{\hbar} (2I+1) B_{1s}
\end{equation}
The vacuum deorientation coefficients for a pure $1s$ configuration
are
\begin{equation}\protect\label{eq:Gkinfty}
G_k^{1s} = 1 - b_k \frac{(\omega \tau)^2}{1+(\omega \tau)^2},
\end{equation}
where $\tau$ is the nuclear lifetime, and
\begin{equation}
b_k = k(k+1)/(2I+1)^2
\end{equation}
Thus for $I=2$ we have $b_2 = 0.24$ and $b_4=0.80$. In the worst
case for $I=2$, $G_2^{1s}({\rm hard core}) = 0.76$, and although
$G_4^{1s}({\rm hard core}) = 0.20$, the $k=4$ term ($a_4$) in the
angular correlation is already small. In fact, for virtually any
reasonable level lifetime (approaching a picosecond or more) and a
$g$-factor value that differs even a little from zero, the hyperfine
field is so big that $G_k^{1s} \approx G_k^{1s}({\rm hard core})$.

\begin{figure}
    \resizebox{0.8\textwidth}{!}{
  \includegraphics[height=.8\textheight]{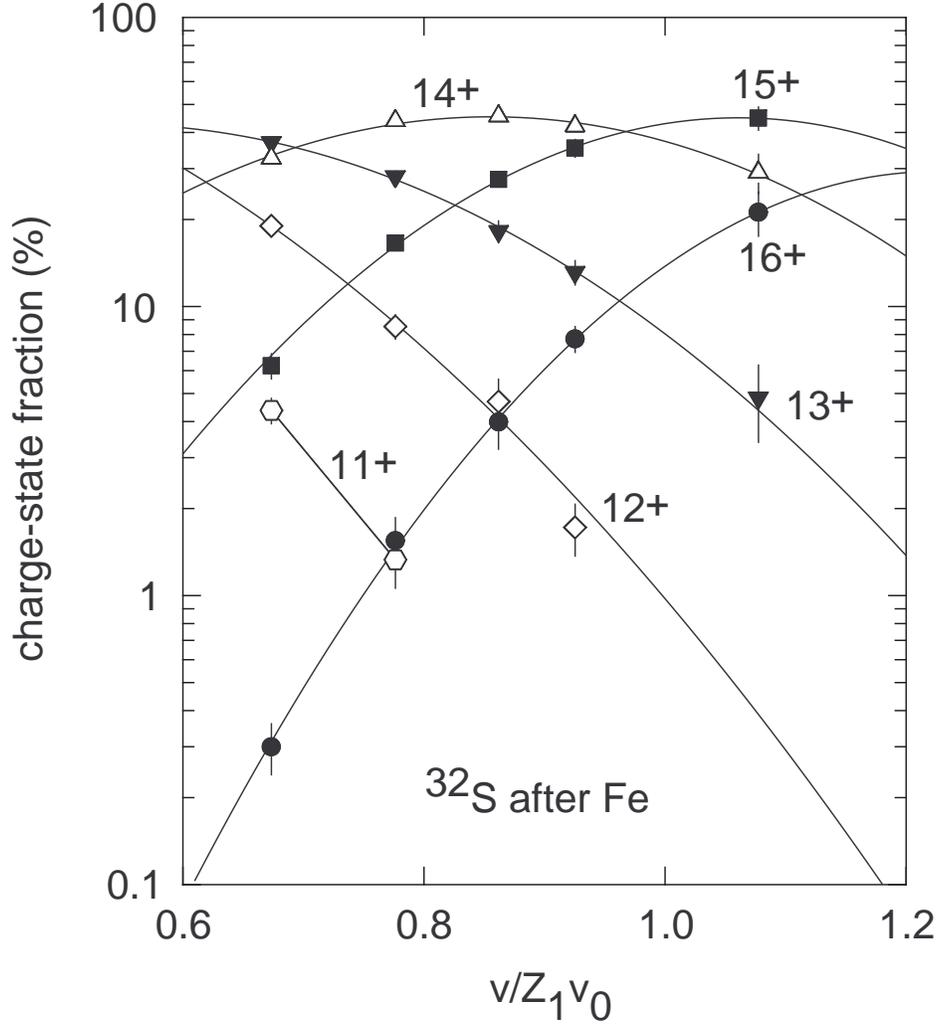}}
\caption{Charge-state fractions for S ions emerging from Fe as a
function of the ion velocity. The highest-velocity point was
obtained with beams boosted by the ANU Linac. The curves are based
on fits to Eq.~(\protect \ref{eq:fitFq}).} \label{fig:fig1}
\end{figure}

We have measured the charge-state fractions for Si and S ions with
energies between about 90 MeV and 240 MeV emerging from Fe and Gd.
(AES, Anna Wilson and Paul Davidson, Nuclear Instruments and Methods
B 243, 265 (2006).) The results for S after Fe are shown in
Fig.~\ref{fig:fig1}. Near $v = Zv_0$ the 15$^+$ charge state is
dominant. The other prolific charge states in their atomic ground
states produce either smaller fields (13$^+$) or no field at all
(14$^+$ and 16$^+$). Under these conditions it is reasonable to
evaluate the average vacuum deorientation coefficients for the
experimental conditions in terms of the fraction of ions that carry
a single electron, $Q_H$. Thus
\begin{equation}
G_k = (1-Q_H) + Q_H G_k^{1s}
\end{equation}
After some simple algebra we get
\begin{equation}
G_k = 1 -Q_H b_k \frac{(\omega \tau)^2}{1+(\omega \tau)^2}.
\end{equation}
Thus provided that $Q_H < 0.5$, $G_2 > 0.88$ and $G_4 > 0.6$.


For S ions with velocities near $0.7Zv_0$ (100 MeV), the Li-like
fields should be included:
\begin{equation}\protect \label{eq:RIVGK}
G_k = 1 -Q_H b_k \frac{(\omega \tau)^2}{1+(\omega \tau)^2} -Q_{Li}
b_k \frac{(\omega \tau/8)^2}{1+(\omega \tau/8)^2}
\end{equation}
where $Q_{Li}$ is the lithium-like fraction, and it is assumed that
\begin{equation}
B_{ns} = 16.7 Z^3/n^3 \;\;\; {\rm Tesla.}
\end{equation}

The experimental data for the charge fractions in
Fig.~\ref{fig:fig1} were fitted to the form
\begin{equation}\label{eq:fitFq}
F(q,v/Z_1v_0) = \exp(a_0 + a_1 v/Z_1v_0 + a_2 (v/Z_1v_0)^2),
\end{equation}
where the parameters $a_0$, $a_1$ and $a_2$ are determined for each
charge state. The fit parameters for the most prolific charge states
are listed in Table~\ref{tab:fitFq}.

\begin{table}
\caption{Results of fits to Eq.~(\protect \ref{eq:fitFq}) for ions
with 3 or fewer electrons.}
\begin{ruledtabular}
\begin{tabular}{cccccccc}
\multicolumn{1}{c}{Beam} & \multicolumn{1}{c}{Target} &
\multicolumn{1}{c}{${q}$}& \multicolumn{2}{c}{$v/Zv_0$ range} &
\multicolumn{3}{c}{Parameters}\\ \cline{4-5} \cline{6-8}
 & & & min & max &
  \multicolumn{1}{c}{$a_0$} &
  \multicolumn{1}{c}{$a_1$} &
  \multicolumn{1}{c}{$a_2$} \\
 \hline
%
%
 Si  & Fe &  14 & 0.83 &   1.26 &   -18.6231&    26.2176& -9.6632\\
  & &  13 & 0.83 &   1.26 &   -9.6665 &    15.3430& -6.5868\\
  & &  12 & 0.83 &   1.26 &   -3.1681 &    7.0924 & -4.8990\\
  & &  11 & 0.83 &   1.26 &   -4.1570 &    9.5851 & -8.0001\\
\\
 S & Fe: &  16 & 0.67 &   1.08 &   -23.6376&    36.5961& -14.9403\\
  &  &  15 & 0.67 &   1.08 &   -14.9891&    26.7520& -12.6089\\
  &  &  14 & 0.67 &   1.08 &   -7.6139 &    15.9538& -9.3268\\
  &  &  13 & 0.67 &   1.08 &   -3.1411 &    8.4963 & -7.8757\\
\\
 S & Gd  &  16 & 0.75 &   1.08 &   -17.8000&    23.8431& -8.6552\\
   &     &  15 & 0.64 &   1.08 &   -12.7697&    20.9555& -9.2347\\
  &  &  14 & 0.64 &   1.08 &   -6.0770 &    11.9903& -6.6953\\
  &  &  13 & 0.64 &   1.08 &   -2.2632 &    5.3773 & -5.1736\\
\\
 Si & Gd &  14 & 0.80 &   1.25 &   -17.9075&    23.8099& -8.4427\\
  & &  13 & 0.80 &   1.25 &   -8.0397 &    11.4628& -4.4435\\
  & &  12 & 0.80 &   1.25 &   -5.6947 &    11.2019& -6.3201\\
  & &  11 & 0.80 &   1.12 &   0.5950  &    0.1204 & -3.0000\\
\\
\end{tabular}
\end{ruledtabular}
\label{tab:fitFq}
\end{table}

The code at present uses the fit to the charge-state results for S
after Fe to interpolate the values of $Q_H$ and $Q_{Li}$ and thus
obtain as realistic an evaluation of $G_k$ as possible. (It will
give a warning if the ion velocity is outside the region where there
are experimental data. The extrapolations must break down because
the charge-state distributions are not truly a gaussian function of
ion velocity.)

The code uses the expression in Eq.~(\ref{eq:RIVGK}) to evaluate the
RIV effect. In the cases considered to date, $\omega \tau$ is so
large that
\begin{equation}
\frac{(\omega \tau)^2}{1+(\omega \tau)^2} \approx  \frac{(\omega
\tau/8)^2}{1+(\omega \tau/8)^2} \approx 1,
\end{equation}
and hence, to an excellent approximation,
\begin{equation}
G_k = 1 - (Q_H + Q_{Li})b_k.
\end{equation}
{\bf Cautionary note:} We know the form of $G(t)$ for the vacuum
deorientation effect, however we measure
\begin{equation}
\langle G_k \rangle = (1/\tau) \int_{t_{exit}}^{\infty} G_k(t)
\exp(-t/\tau) dt,
\end{equation}
where $t_{exit}$ is the time at which the ions emerge into vacuum.
The code at present assumes $t_{exit} = 0$ in the evaluation of the
$\langle G_k \rangle $, as is the case for Eq.~(\ref{eq:Gkinfty}).
This gives an accurate result if $t_{exit} \ll \tau$. Thus the
present calculations for $^{40}$S ($\tau \sim 20$~ps) should be
fine, while those for $^{38}$S ($\tau \sim 5$~ps) might tend to over
estimate the vacuum deorientation effect slightly. A more rigorous
evaluation of the $\langle G_k \rangle $ value applicable in the
experiments is available from the Monte Carlo Doppler Broadened Line
Shape (DBLS) code, GKINT\_MCDBLS.


\section{Some pedagogical plots of the perturbed correlations}

\subsection{TF precessions in the nuclear (projectile) frame}

While the effect of the precession is most pronounced for detectors
in the horizontal plane, the other planes are also expected to show
useful precession effects. The following figures \ref{fig:fig4} -
\ref{fig:fig6} illustrate how the projectile (or nuclear frame)
precession varies with the phi angle of the $\gamma$-ray detector.

\begin{figure}
    \resizebox{0.5\textwidth}{!}{
  \includegraphics[height=.8\textheight]{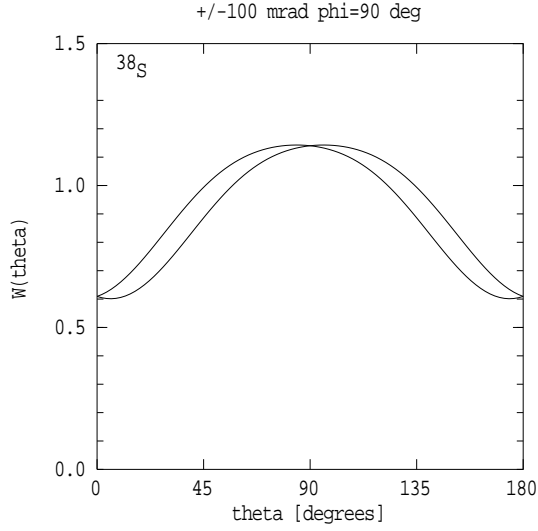}}
\caption{Perturbed angular correlations in the CM frame as observed
in the horizontal, or $\phi=90^\circ$, plane.} \label{fig:fig4}
\end{figure}

\begin{figure}
    \resizebox{0.5\textwidth}{!}{
  \includegraphics[height=.8\textheight]{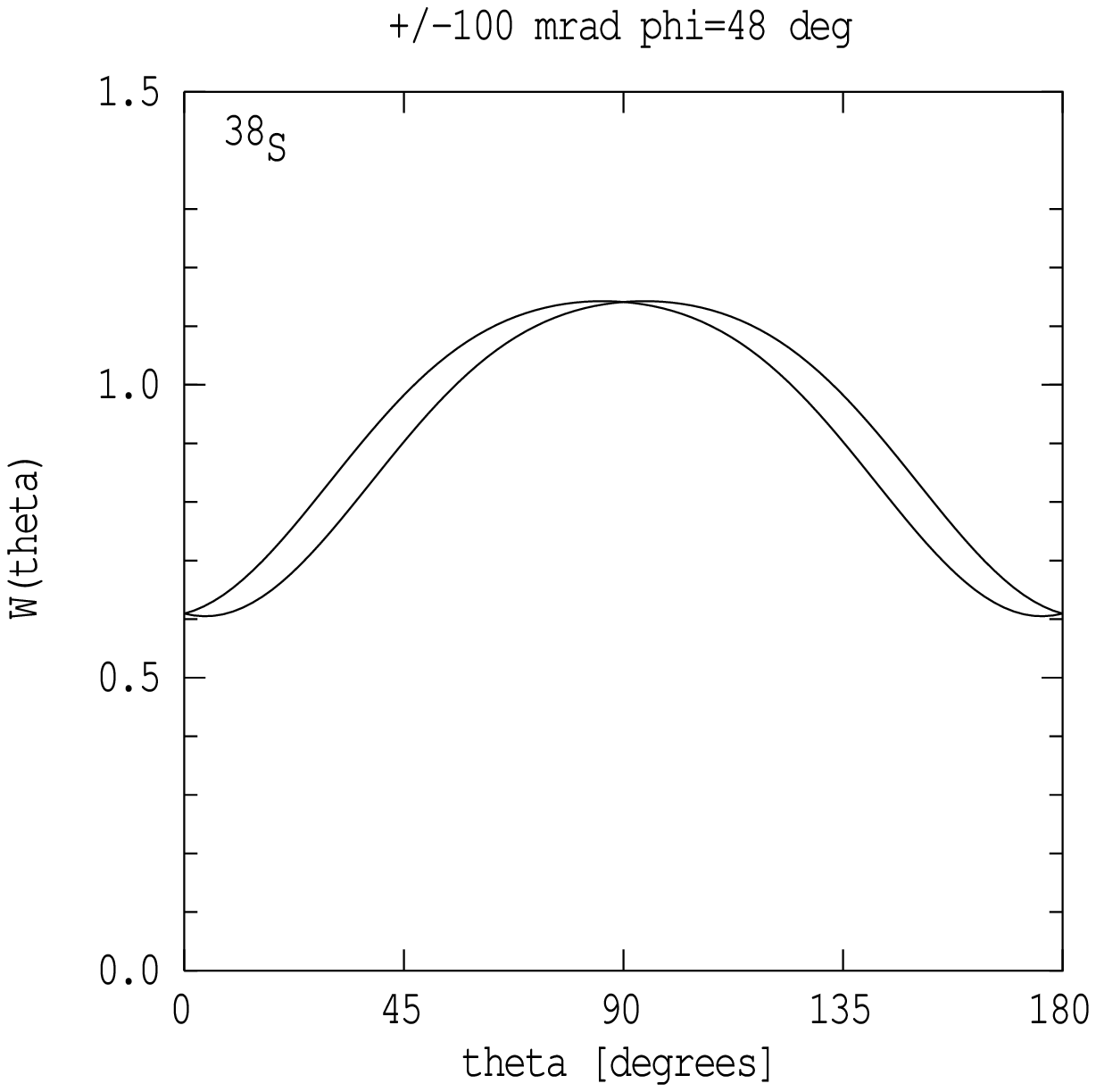}}
\caption{Perturbed angular correlations in the CM frame as observed
in the $\phi=\pm 48^\circ$ or $\phi=\pm 228^\circ$ planes.}
\label{fig:fig5}
\end{figure}

\begin{figure}
    \resizebox{0.5\textwidth}{!}{
  \includegraphics[height=.8\textheight]{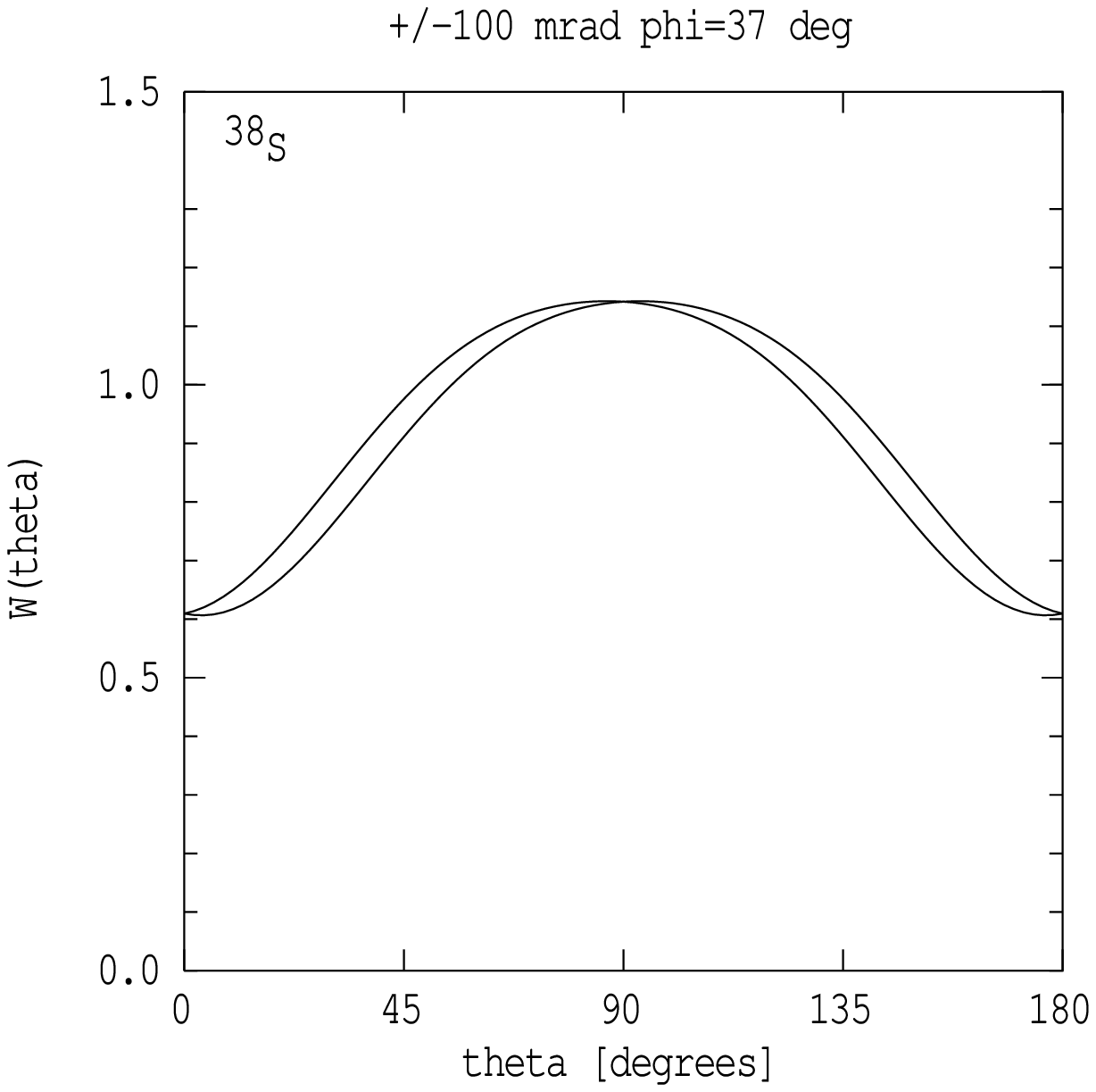}}
\caption{Perturbed angular correlations in the CM frame as observed
in the $\phi=\pm 37^\circ$ or $\phi=\pm 217^\circ$ planes.}
\label{fig:fig6}
\end{figure}

\subsection{RIV perturbation in the lab frame}

\begin{figure}
    \resizebox{0.8\textwidth}{!}{
  \includegraphics[height=.8\textheight]{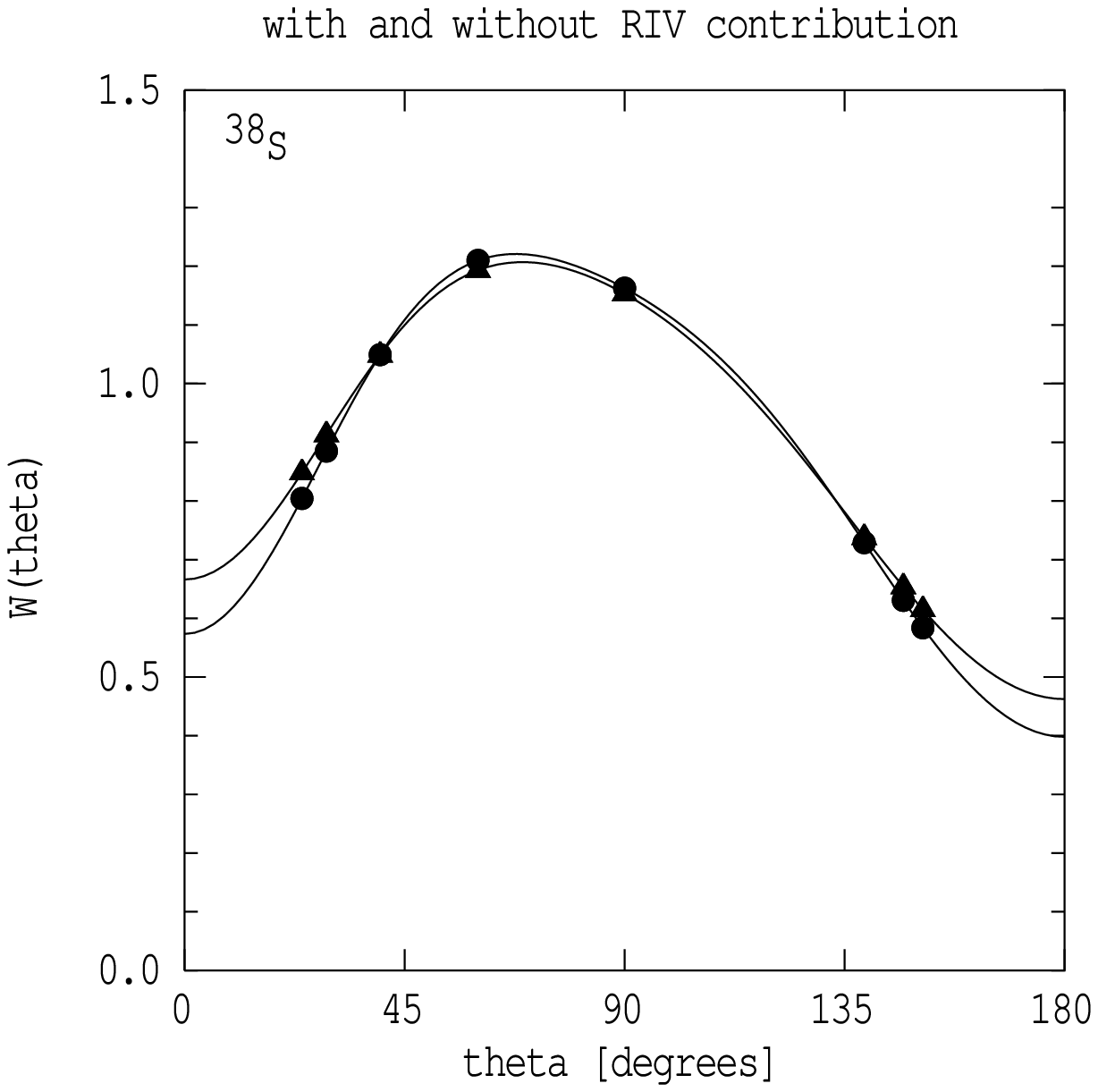}}
\caption{Angular correlation in the lab frame with and without the
RIV contribution.} \label{fig:rivlab}
\end{figure}

Fig.~\ref{fig:rivlab} shows the angular correlation in the lab frame
with and without the RIV contribution. In this case the particle
scattering angle is between 2.5$^\circ$ and 5.5$^\circ$. The
$^{38}$S beam energy is 1540 MeV. The unattenuated angular
correlation coefficients are $a_2 = -0.4233$ and $a_4 = -0.0990$.
For a H-like fraction of   $Q_H = 0.1269$ and a Li-like fraction of
$Q_{Li} = 0.3108$, $G_2 = 0.8978$ and $G_4 = 0.6595$. The Lorentz
effects are evaluated using $\beta = 0.089$.

Note that the RIV effect on the angular correlation is modest and
that it is very difficult to measure the RIV attenuation because
$\gamma$-ray detectors cannot be placed at either 0$^\circ$ or
180$^\circ$ to the beam in intermediate-energy Coulomb excitation
measurements. The measurements of the charge-state fractions are
therefore very important.

\begin{table}[ht]
\caption{Additional details of the angular correlation in
Fig~\protect\ref{fig:rivlab} for the case without vacuum
deorientation.}
\begin{ruledtabular}
\begin{tabular}{rrrrrrr}
 posn & $\theta$ & $\phi$  &  $d\Omega_{\rm nuc}/d\Omega_{\rm lab} $ &
 $W(\theta)_{\rm lab}$ &     $S(\theta)$   &   $S^2W$  \\ \hline
%
   1  &  24.0 &  180.0 &  1.1763 &  0.8013 &  0.0000 &  0.000 \\
   2  &  40.0 &  131.0 &  1.1433 &  1.0459 &  0.5937 &  0.369 \\
   3  &  90.0 &  112.0 &  0.9920 &  1.1630 & -0.0381 &  0.002 \\
   4  & 147.0 &  143.0 &  0.8584 &  0.6329 & -0.6467 &  0.265 \\
   5  & 139.0 &  228.0 &  0.8705 &  0.7315 &  0.6896 &  0.348 \\
   6  &  60.0 &  241.0 &  1.0870 &  1.2077 & -0.2589 &  0.081 \\
   7  & 151.0 &   90.0 &  0.8533 &  0.5861 & -1.1148 &  0.728 \\
   8  & 151.0 &  270.0 &  0.8533 &  0.5861 &  1.1148 &  0.728 \\
   9  &  29.0 &   91.0 &  1.1675 &  0.8816 &  1.0543 &  0.980 \\
  14  &  29.0 &  271.0 &  1.1675 &  0.8816 & -1.0543 &  0.980 \\
  17  &  24.0 &    1.3 &  1.1763 &  0.8013 &  0.0254 &  0.001 \\
  18  &  60.0 &   61.0 &  1.0870 &  1.2077 &  0.2589 &  0.081 \\
  19  & 139.0 &   46.0 &  0.8705 &  0.7315 & -0.6675 &  0.326 \\
  20  & 147.0 &  323.0 &  0.8584 &  0.6329 &  0.6467 &  0.265 \\
  21  &  90.0 &  292.0 &  0.9920 &  1.1630 &  0.0381 &  0.002 \\
  22  &  40.0 &  310.0 &  1.1433 &  1.0459 & -0.6027 &  0.380 \\
\end{tabular}
\end{ruledtabular}
\end{table}

\begin{table}[ht]
\caption{Additional details of the RIV attenuated angular
correlation in Fig~\protect\ref{fig:rivlab}.}
\begin{ruledtabular}
\begin{tabular}{rrrrrrr}
 posn & $\theta$ & $\phi$  &  $d\Omega_{\rm nuc}/d\Omega_{\rm lab} $ &
 $W(\theta)_{\rm lab}$ &     $S(\theta)$   &   $S^2W$  \\ \hline
 \\
   1  &  24.0 &  180.0 &  1.1763 &  0.8450 &  0.0000 &  0.000 \\
   2  &  40.0 &  131.0 &  1.1433 &  1.0454 &  0.5165 &  0.279 \\
   3  &  90.0 &  112.0 &  0.9920 &  1.1536 & -0.0468 &  0.003 \\
   4  & 147.0 &  143.0 &  0.8584 &  0.6562 & -0.5152 &  0.174 \\
   5  & 139.0 &  228.0 &  0.8705 &  0.7404 &  0.5773 &  0.247 \\
   6  &  60.0 &  241.0 &  1.0870 &  1.1911 & -0.2669 &  0.085 \\
   7  & 151.0 &   90.0 &  0.8533 &  0.6169 & -0.8662 &  0.463 \\
   8  & 151.0 &  270.0 &  0.8533 &  0.6169 &  0.8662 &  0.463 \\
   9  &  29.0 &   91.0 &  1.1675 &  0.9096 &  0.8472 &  0.653 \\
  14  &  29.0 &  271.0 &  1.1675 &  0.9096 & -0.8472 &  0.653 \\
  17  &  24.0 &    1.3 &  1.1763 &  0.8450 &  0.0196 &  0.000 \\
  18  &  60.0 &   61.0 &  1.0870 &  1.1911 &  0.2669 &  0.085 \\
  19  & 139.0 &   46.0 &  0.8705 &  0.7404 & -0.5588 &  0.231 \\
  20  & 147.0 &  323.0 &  0.8584 &  0.6562 &  0.5152 &  0.174 \\
  21  &  90.0 &  292.0 &  0.9920 &  1.1536 &  0.0468 &  0.003 \\
  22  &  40.0 &  310.0 &  1.1433 &  1.0454 & -0.5242 &  0.287 \\
\end{tabular}
\end{ruledtabular}
\end{table}

\section{Figure of merit: optimizing experimental conditions}

GKINT provides the quantities designated $S^2W$, $S^2N$ and
$S^2\phi^2N$ evaluated at the SeGA detector angles as figures of
merit. This section explains the rationale of these choices as
figures of merit.

We begin by obtaining an estimate of the experimental error on the
measured precession for a given number of accumulated counts in a
single pair of detectors. Since the precession angles are relatively
small, we can assume that
\begin{equation}
N(+\theta_\gamma,\uparrow) \approx N({-\theta_\gamma},\downarrow)
\approx N({+\theta_\gamma},\downarrow) \approx
N({-\theta_\gamma},\uparrow) \approx N .
\end{equation}
Thus $N$ is the total number of counts recorded in {\em one}
detector for {\em one} field direction. With the further
approximation that
\begin{equation}
\sigma_N \approx \sqrt{N},
\end{equation}
it follows that
\begin{displaymath}
\sigma_{\Delta\theta} \approx \frac{1}{2 S \sqrt{N}}.
\end{displaymath}
It is usually more convenient to work with a figure of merit that
scales with the number of counts, and hence is inversely
proportional to the required beam time. Thus we can calculate
quantities proportional to $ S^2 N$ as the figure of merit and note
that we must maximize it for optimum sensitivity.

GKINT calculates both $S^2W$ and $S^2 W \sigma $, denoting the
latter as $S^2N$. Generally $S^2N = S^2 W \sigma $ will be the
quantity of interest. ($S^2W$ is retained in the output to give an
indication of the angular correlation contribution, separate from
the cross section.) These figures of merit are independent of the
magnitude of $\Delta \theta$. Clearly, in a real experiment the goal
is to minimize $\sigma_{\Delta\theta}/\Delta\theta$. As another
figure of merit GKINT also evaluates
\begin{equation}
S^2 \phi^2 W \sigma e^{(-T_{\rm exit}/\tau)},
\end{equation}
which is proportional to $(\sigma_{\Delta\theta}/\Delta\theta)^2$
for the fraction of excited nuclei that experience the full
transient-field precession and decay in vacuum beyond the target.
Here $T_{\rm exit}$ is the average time at which beam ions exit from
the target.

The calculations are given with and without the effect of the RIV
attenuation. Since the RIV effect can not be avoided, the
unattenuated angular correlations are given mainly as a point of
reference.

\section{Effect of a mask on the particle detector}

Table \ref{tab:mask} summarizes some calculations made to assess the
use of different masks on the particle detector. The main benefit of
the mask is to reduce the Rutherford rate. The small reduction in
Coulex count rate is compensated for by an increase in the
anisotropy of the angular correlation.

\begin{table}[ht]
\caption{Effect of a mask on the particle detector. The calculations
are for a standard $^{38}$S configuration with $E_{\rm beam} =
1540$~MeV. The target is 355 mg/cm$^2$ Au followed by 100 mg/cm$^2$.
The stopping powers are scaled by a factor of 1.02.}
\begin{ruledtabular}
\begin{tabular}{cccc}
$\theta_p^{\rm min}$ & \multicolumn{2}{c}{Count rate/1000 beam ions
(Hz)} & $S^2W \sigma$ \\ \cline{2-3}
 & Coulex & Rutherford &\\
\\ \hline
\multicolumn{4}{l}{$\theta_p^{\rm max} = 5^\circ$} \\
0   &   0.0596   &   1000   & 762 \\
1   &   0.0598   &    290   & 778 \\
2   &   0.0523   &     63   & 856 \\
2.5 &   0.0466   &     36   & 869 \\
3   &   0.0395   &     21   & 836 \\
\\
\multicolumn{4}{l}{$\theta_p^{\rm max} = 5.5^\circ$} \\
0   &   0.0720   &   1000   & 1119 \\
1   &   0.0713   &    292   & 1138 \\
2   &   0.0647   &     66   & 1236 \\
2.5 &   0.0589   &     38   & 1260 \\
3   &   0.0518   &     24   & 1237 \\
\end{tabular}
\end{ruledtabular}
\label{tab:mask}
\end{table}

\end{document}